\documentclass[aps,twocolumn,secnumarabic,balancelastpage,amsmath,amssymb,nofootinbib,floatfix]{revtex4-1}

\usepackage{graphicx}   
\usepackage{bm}      
\usepackage{subcaption}
\usepackage[colorlinks=true]{hyperref} 
\usepackage{braket}
\usepackage{array}

\begin{document}


\title{Semi-Transparent Solar Cell enabled by Frequency Selective Light Trapping}

\author{Duncan C. Wheeler\textit{$^{1}$}, Yichen Shen$^{\ast}$\textit{$^{1}$}, Yi Yang\textit{$^{12}$}, Svetlana V. Boriskina\textit{$^{3}$}, Yi Huang\textit{$^{3}$}, Ognjen Ilic\textit{$^{4}$}, Gang Chen\textit{$^{3}$}, Marin Solja{\v{c}}i{\'c} \textit{$^{1}$}}

\begin{abstract}
We propose a frequency selective light trapping scheme that enables the creation of more visually-transparent and yet simultaneously more efficient semitransparent solar cells. A nanoparticle scattering layer and photonic stack back reflector create a selective trapping effect by total internal reflection within a medium, increasing absorption of IR light. We propose a strong frequency selective scattering layer using spherical $\textrm{TiO}_{2}$ nanoparticles with radius of 255 nm and area density of 1.1\% in a medium with index of refraction of 1.5. Using detailed numerical simulations for this configuration, we find that it is possible to create a semitransparent silicon solar cell that has a Shockley Queisser efficiency of $12.0\%\pm0.4\%$ with a visible transparency of $60.2\%\pm1.3\%$, $13.3\%\pm1.3\%$ more visibly-transparent than a bare silicon cell at the same efficiency.
\end{abstract}

\maketitle


\footnotetext{Department of Physics, MIT, USA; E-mail: duncanw@mit.edu, ycshen@mit.edu}
\footnotetext{Department of Electrical Engineering and Computer Science, MIT, USA}
\footnotetext{Department of Mechanical Engineering, MIT, USA}
\footnotetext{Department of Applied Physics, Caltech, USA}


As the world experiences growing demand for energy, the cost of non-renewable energy, both financially and environmentally, becomes increasingly apparent \cite{EnergyData}. Consequently, new techniques in renewables have garnered increased attention. Recent studies have examined building-integrated photovoltaics as one way to increase renewable energy generation in urban environments \cite{BIPV}. Semitransparent solar cells are essential to the success of building-integrated photovoltaics, as they can be installed in window locations to generate energy while preserving room lighting on building sides. However, semitransparent solar cells typically have low efficiency; they must become more efficient and transparent before they can become widely adopted. In this paper, we propose a frequency selective light trapping scheme that enables improved performance of semitransparent solar cells \cite{into-source}.


\section{Background}
\label{sec: Background}

\begin{figure*}[t]
	\includegraphics[width = \textwidth]{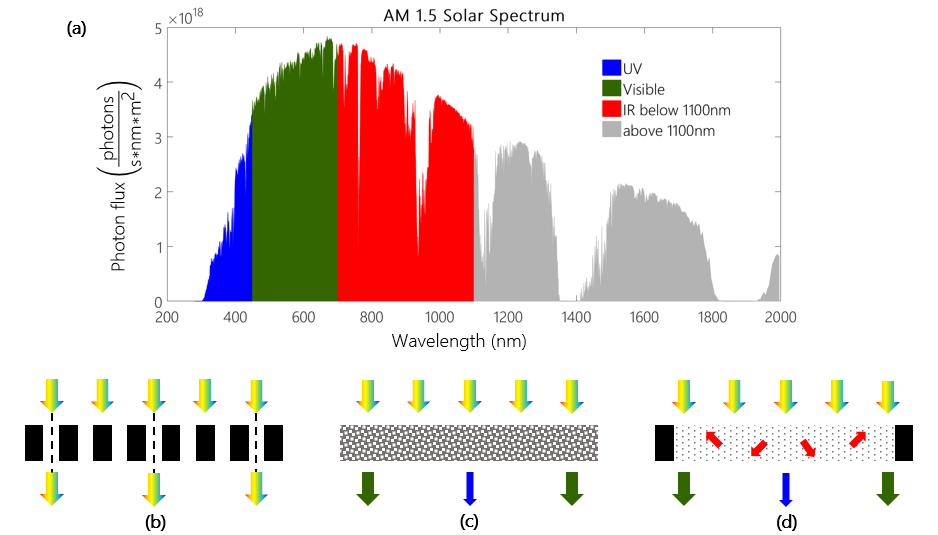}
	\caption{Existing transparent solar cell technology. (a) Incoming photon flux from the sun divided into UV, Visible, and IR light above the silicon bandgap. (b) A semitransparent cell made from an etched standard cell. (c) A semitransparent cell with a frequency selective active layer. (d) A semitransparent luminescent solar concentrator which guides down-converted light to standard solar cells}
	\label{fig:TheorySummary}
\end{figure*}

Designing semitransparent solar cells leads to an inherent trade-off between transparency and efficiency, as transmitted light cannot be converted to energy. However, human eyes are only sensitive to a small range of the electromagnetic spectrum. Acknowledging this, we work around the efficiency-transparency trade-off. As Figure~\ref{fig:TheorySummary}\color{red}a \color{black}shows, over half of the incoming photons with energies above the silicon bandgap lie outside the visible spectrum, and mostly in the IR region. While photons in the visible spectrum are more energetic, standard semiconductor solar cells generate one carrier for each absorbed photon above the bandgap, regardless of frequency. Therefore, by tailoring the absorption spectrum of a standard semiconductor solar cell, such that it absorbs all incoming IR light above the bandgap energy while transmitting all visible light, we can theoretically achieve a 100\% visibly transparent solar cell with high efficiencies. A more detailed calculation of this theoretical limit shows that the Shockley Queisser efficiency limit for such a transparent solar cell is 21\%, almost two thirds that of normal single junction semiconductor solar cells \cite{EfficiencyLimit}.


While semitransparent solar cells could, in principle, reach reasonably high efficiencies, existing technologies are limited by multiple constraints. The most efficient current strategy, seen in Figure~\ref{fig:TheorySummary}\color{red}b\color{black}, reaches up to 10\% efficiency, but remains relatively opaque. These cells consist of traditional semiconductor solar cells with gaps etched into their surfaces. Such a layout produces a screen-porch like effect where enough light passes through that human eyes can see an image, but the resulting image is blurred and darkened, making the presence of a solar cell obvious. Because these cells fail to distinguish between visible and non-visible light, they truly are limited by a trade-off between transparency and efficiency.

Other semitransparent solar cells include organic solar cells, quantum dot solar cells, and transparent luminescent solar concentrators (TLSC). Figure~\ref{fig:TheorySummary}\color{red}c\color{black}~shows a cell where the active material itself is designed to absorb non-visible light only, such as a semitransparent organic cell or quantum dot cell. Figure~\ref{fig:TheorySummary}\color{red}d\color{black}~illustrates a TLSC where molecular dyes absorb incoming non-visible light and re-emit the energy at a shifted frequency, which is then directed by total internal reflection in a waveguide to a standard solar cell. Because these strategies use materials that intrinsically absorb non-visible light, they achieve very high transparencies, as high as 84\% in the case of TLSC's \cite{TransparentLSC1}. However, each of these strategies face important limitations that reduce their efficiencies to less than 4\% \cite{TransparentLSC1,TransparentLSC2,TransparentLSC3,TransparentOrganic1,TransparentOrganic2,TransparentQD1}. Organic cells suffer from limited carrier mobility, reducing their overall efficiency \cite{OrganicTheory}. Quantum Dot cells suffer from incomplete absorption of desired light in addition to poor carrier mobility \cite{PhotovoltaicLimits}.TLSC's can in principle reach high efficiencies; however, surface losses in the waveguide and re-absorption losses in the dyes lead to efficiency losses that are difficult to avoid \cite{TransparentLSCTheory}.

\section{Concept}

In this paper, we propose an approach to improve semitransparent solar cells through frequency selective light trapping. The basic design consists of a semitransparent active layer placed between a frequency selective scattering layer on top and a frequency selective reflective layer on bottom. The scattering and reflective layers trap the IR light, leading to an increased path length through the active layer of the solar cell between them. At the same time, visible light passes through the frequency selective layers and semitransparent active layer largely unaffected. Such a system increases the efficiency of semitransparent solar cells without decreasing their transparencies. Conversely, we can use more transparent solar cells without loss of efficiency, allowing us to explore ultra-thin semiconductors as semitransparent active layers with strong carrier properties.

In principle, our light trapping scheme works similar to those used for current ultra-thin photovoltaics. While many different light trapping techniques exist, one common method relies on randomly textured surfaces that scatter light isotropically \cite{Sveta1, Sveta2}. When light isotropically scatters inside a medium, internal reflection traps most photons inside the medium, increasing light concentration. This is the basis for the Yablonovitch limit on light trapping and for our method of frequency selective light trapping \cite{Yablonovitch}. Using a layer of nanoparticles designed to isotropically scatter IR light, we trap such light through total internal reflection. By using a frequency selective mirror on the back surface of our system, we reduce escape cone losses to only the top surface.

\begin{figure}[h]
	\centering
	\includegraphics[width = \linewidth]{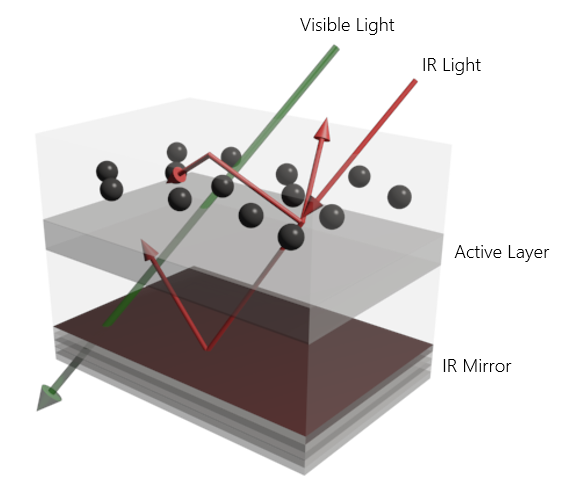}
	\caption{Model for semitransparent solar cell using frequency selective light trapping. A very thin active layer absorbs little light. Nanoparticles and a 1-D photonic crystal create a frequency light trapping effect, increasing IR path lengths through the active layer and resulting in stronger absorption.}
	\label{fig:CellDesign}
\end{figure}

\subsection{Active Layer}

To evaluate our design, we first consider the active layer. This layer generates the energy in our setup and can be any semitransparent solar cell. We consider two main options for the active layer, first of which are ultra-thin semiconductors. Using the absorption coefficient of silicon along with the Beer-Lambert law, we find that a silicon layer with a thickness on the order of 800nm or less is needed for a visible transparency of 60\% or more \cite{SiliconData}. Similarly, a Gallium Arsenide (GaAs) layer of 75nm thickness or less gives a visible transparency of 60\% or more \cite{GaAsData}. By comparing these two materials along with other semiconductors, we find that silicon's low absorption allows it to more easily be made visibly transparent than the other active layer materials we consider. In addition, silicon's absorption coefficient remains relatively high well into the IR spectrum, enabling more effective absorption of non-visible light.


The second active layer option we consider are quantum dots. While quantum dot solar cells face low efficiencies regardless of their transparency, they are relatively easy to make initially transparent and are therefore useful to consider for future experiments to verify the effectiveness of frequency selective light trapping. Specifically, we use data for tetrabutylammonium iodide (TBAI)-exchanged PbS films to model the active layer \cite{quantum}. While quantum dots solar cells do not form a flat sheet like semiconductor solar cells, we model the active layer as a sheet of PbS-TBAI with a set thickness to determine the amount of active layer material that would result in a high transparency.  Again, by using active layer absorption coefficients with the Beer-Lambert Law, we find that a PbS-TBAI thickness of around 45nm provides a visible transparency of 60\%.


\subsection{Frequency Selective Reflector}

After selecting active layer compositions, we consider the reflecting layer. For our design, we aim for a perfect frequency selective mirror that transmits all visible light while reflecting all IR light. Ilic et al. designed and fabricated a thin-film multilayer stack with 2 materials to reflect IR light while transmitting visible light for the purpose of improving the efficiency of incandescent light sources. This structure reached 92\% reflectance over a wide range of angles and frequencies in the IR spectrum \cite{Ognjen}. Ilic et al. also showed the possibility for making near ideal frequency selective reflectors using more complicated structures. Therefore, we assume a perfect frequency selective reflector in the calculations that follow.\footnote[5]{Possible behavior of a non-ideal reflector can be found in Appendix \ref{app:reflector}}  Using a perfect reflector, we then consider the scattering layer, which creates the IR light trapping effect.


\section{Scattering Layer Optimization}

\begin{figure*}[t]
	\centering
	\includegraphics[width = \linewidth]{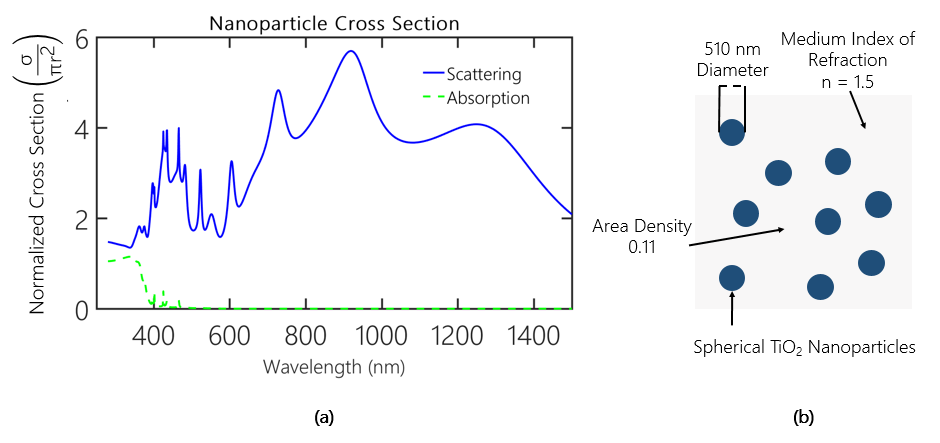}
	\caption{Results of single layer nanoparticle optimization. (a) The scattering and absorption cross sections of a single $\textrm{TiO}_2$ nanoparticle of 255 nm radius, with a broad IR resonance peak. (b) Diagram of optimized scattering layer with each optimization parameter labeled.}
	\label{fig:OptimizationParameters}
\end{figure*}

To determine the best design of the scattering nanoparticles, we perform a global optimization using the Multi-Level Single-Linkage (MLSL) algorithm running a BOBYQA local optimization algorithm, both implemented by NLopt, a free nonlinear optimization library \cite{NLopt,NLopt_MLSL,NLopt_BOBYQA}.  We consider four main parameters that characterize the layer. As shown in Figure~\ref{fig:OptimizationParameters}\color{red}b\color{black}, these parameters are nanoparticle material index, radius (r), and area density (n) in addition to medium index. The material and radius determine how each nanoparticle scatters light, and are primarily responsible for creating the resonance we aim for. The area density ($n=$ cell area / [$\pi*r^2$ $*$ \# of particles]) determines nanoparticle separation and how the scattering layer acts overall given the behavior of individual particles.   We assume that the nanoparticles are deposited in a single layer, and will therefore only scatter a light ray once each time the ray passes through the scattering layer.  Finally, the medium index controls the escape angle as well as the index contrast between the medium and the nanoparticles, affecting both light scattering and light trapping. To account for all parameters, we perform our optimization several times over nanoparticle radius and area density while manually varying the nanoparticle material and medium index between each run.  For nanoparticle material, we consider $\textrm{Ta}_2\textrm{O}_5$, silver, gold, and $\textrm{TiO}_2$. We then choose the material and medium index that give the best optimized figure of merit (FOM).


\subsection{Defining a Figure of Merit}

To quantify the performance of our light trapping scheme, we use a figure of merit that characterizes the system as one which allows us to use a more transparent cell without losing efficiency. Our optimization figure of merit can be written as below:
\begin{equation}
	\textrm{FOM} = 60\% - T_{\textrm{bare}}.
	\label{eq:FOM}
\end{equation}
$T_{bare}$ represents the transparency of a bare active layer when it achieves the same efficiency that our frequency selective light trapping system achieves with a transparency of 60\%. Because we define our figure of merit this way, we must calculate the efficiency and transparency of our system at different active layer thicknesses.

To find these efficiencies and transparencies, we first find the scattering and absorption cross section of individual nanoparticles. Using Mie Scattering Theory, we calculate the scattering of a plane wave expressed in terms of the spherical harmonics with angular momentum number up to 15th order to account for all dominant terms and determine the nanoparticle cross section \cite{scatteringBook}. We then use the Beer-Lambert law to convert the scattering and absorbtion cross sections ($\sigma_{s}$ and $\sigma_{a}$), along with particle area density ($n$), into normalized scattering and absorption probabilities ($P_{s}$ and $P_{a}$) for the whole scattering layer as follows:
\begin{equation}
	P_{s} = \frac{1-e^{-n(\sigma_{s}+\sigma_{a})}}{1+\frac{1-e^{-n\sigma_{a}}}{1-e^{-n\sigma_{s}}}}
    \quad \textrm{and} \quad 
    P_{a} = \frac{1-e^{-n(\sigma_{s}+\sigma_{a})}}{1+\frac{1-e^{-n\sigma_{s}}}{1-e^{-n\sigma_{a}}}}.
    \label{eq:probability}
\end{equation}
This equation assumes a single layer of nanoparticles and therefore only considers a single possible scattering or absorption event each time a light ray passes through the scattering layer.  By limiting the area density to less than 30\%, we increase the likelihood of having independent scattering events.  After our optimization, we then double check to ensure that the cross section areas of the optimized nanoparticles do not overlap, ensuring that the scattering events are independent.


\subsection{Ray Trace to Find Absorbed Photon Flux}

Having characterized the scattering layer, we perform a Monte Carlo ray tracing calculation to determine our final FOM. We consider a set of photons entering the top of our solar cell at normal incidence. When a photon passes through the scattering layer, it scatters or is absorbed with probabilities obtained from our Mie scattering calculations above. If the photon scatters, we assume isotropic scattering, becuase the light will scatter off a single nanoparticle without interacting with other nanoparticles in the scattering layer.  As a reult, we assign a random direction to the scattered photon. While our ray tracing is 3 dimensional, only the angle of the ray from normal incidence effects our result.  Therefore, we convert the isotropic solid angle distribution to a single angle, defined from vertical, using a sine distribution, making our ray tracing 2 dimensional. We generate a sine distribution by using equation~\eqref{eq:angle} to convert a random number between 0 and 1, R, to the scattering angle, $\theta$. 
\begin{equation}
	\theta=\arccos(1-2*R)
    \label{eq:angle}
\end{equation}
When a photon passes through the active layer, we determine the path length of the light through the active by dividing the active layer thickness, T, by the sine of the angle from horizontal, as in equation~\eqref{eq:pathlength}.
\begin{equation}
	\textrm{Path Length}=\frac{T}{\sin(\theta-\frac{\pi}{2})}
	\label{eq:pathlength}
\end{equation}
We sum these path lengths for individual photons to track the total path length of each photon through the active layer for later calculations. When a photon encounters the reflective layer at the bottom of the cell, we check for total internal reflection for visible wavelengths and assume perfect reflection for all other wavelengths. Similarly, when a photon hits the top of the cell, we check for total internal reflection for all wavelengths. If a photon is reflected, we flip the direction by setting the new angle to $\pi$ minus the old angle. The ray trace continues for each photon with an additional scattering every time the photon interacts with the scattering layer until that photon is absorbed by the scattering layer or escapes from either surface. 

To evaluate our FOM, we perform the ray trace described above for 1000 photons each over vacuum wavelengths spaced by 10nm starting at 280nm and continuing to 1450nm, which is above the silicon bandgap. From the total path lengths calculated, we determine the average path length of each wavelength through the active layer. By using the Beer-Lambert law we then determine the absorption of our system for each wavelength. Integrating the absorption over the solar spectrum, we calculate the total absorbed incoming photon flux \cite{solarSpectrum}. Similarly, by tracking the fraction of visible photons that escape through the back surface, we multiply this fraction by the non-absorbed photon flux in the visible spectrum to obtain the visible transparency of our system.  Some of the light accounted for in this transparency was scattered by the scattering layer and might lead to a blurred image, resulting in the solar cell being just translucent rather than transparent \cite{translucent1, translucent2}.  When we only consider non-scattered light and calculate in-line transparency, we find a decrease of roughly 3 percentage points, from 60\% transparency to 57\% in-line transparency, leading to a roughly 5\% haze.  This is similar to the effects observed in some attempts to create tunable windows \cite{translucent2}.

\begin{figure*}[t]
	\centering
	\includegraphics[width = \linewidth]{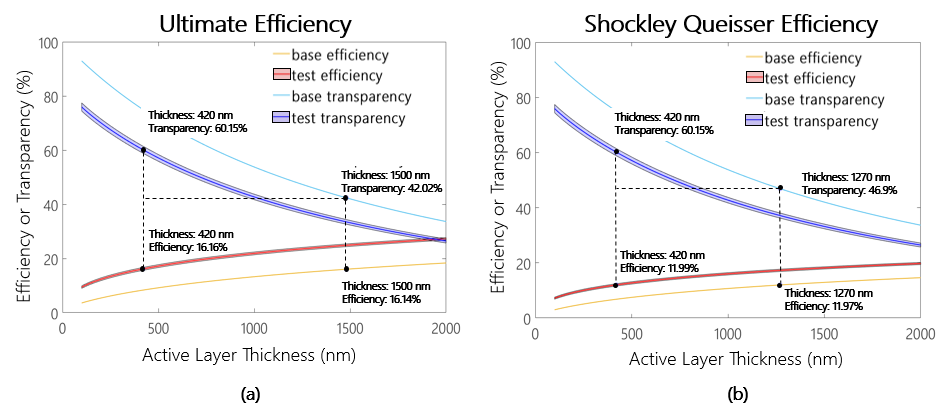}
	\caption{Improvement in transparency at same efficiency due to frequency selective light trapping, (a) using ultimate efficiency and (b) using Shockley Queisser efficiency}
	\label{fig:OptimizationResults}
\end{figure*}


To simplify our calculations we make two approximations. First, the parameters that effect our ray trace are the scattering and absorption probabilities of the scattering layer along with the refraction index of the medium. Therefore, rather than performing the ray trace over each wavelength for each nanoparticle radius and area density, we perform the ray trace for a set of absorption and scattering probabilities in two cases: visible and non-visible wavelengths. We then interpolate the results to quickly access the behavior of our system given the absorption and scattering probability determined by our Mie scattering calculations. Second, the path length of a photon through the active layer scales linearly with the active layer thickness. Therefore, rather than repeating a ray tracing calculation for new active layer thicknesses, we multiply the average path lengths of our original ray tracing calculation by the ratio of active layer thicknesses to obtain new average path lengths and quickly determine new absorptions and transparencies.

\subsection{Efficiency Calculation and Final FOM}
We calculate two different efficiencies from our absorbed photon flux, ultimate efficiency and Shockley Queisser efficiency. Ultimate efficiency assumes that every absorbed photon is converted perfectly into the bandgap energy. To calculate ultimate efficiency, we multiply the absorbed photon flux by the bandgap energy of silicon and divide by the total integrated power of the AM 1.5 solar spectrum \cite{solarSpectrum}. Shockley Queisser efficiency provides an efficiency limit accounting for radiative recombination, more accurately reflecting real world behavior \cite{ShockleyOriginal}. We base our Shockley Queisser calculations off those performed by R$\ddot{\textrm{u}}$hle, which perform a discrete calculation and can use interpolated absorption data from our Monte Carlo simulation \cite{ShockleyCalc}. The original Shockley Queisser calculation assumed an absorption of 1 for frequencies above the band gap and 0 for frequencies below the bandgap.  We also assume an absorption of 0 below the silicon bandgap, but we interpolate absorptions calculated through the ray tracing above to obtain non-unity absorptions at frequencies above the bandgap.


With the efficiency and transparency of our system as a function of thickness, we then evaluate our FOM, as in equation~\eqref{eq:FOM}. Figure~\ref{fig:OptimizationResults} visualizes this FOM by plotting the calculated transparencies and efficiencies of our system against active layer thickness along with those values for a bare active layer, which gives the base efficiencies. We determine the thickness at which our system reaches a transparency of 60\%, find the corresponding efficiency, determine the thickness at which a bare active layer has the same efficiency, and compare the resulting transparency with our system's 60\% transparency.

As seen in Figure~\ref{fig:OptimizationResults}, different efficiency calculations give different FOM results. Figure~\ref{fig:OptimizationResults}\color{red}a\color{black}~shows a transparency improvement of $18.2\pm1.3$ percentage points, from $42.0\%$ to $60.2\%$, when using ultimate efficiency, while Figure~\ref{fig:OptimizationResults}\color{red}b\color{black}~shows a transparency improvement of $13.3\pm1.3$ percentage point, from $46.9\%$ to $60.2\%$, when using Shockley Queisser efficiency. This reduction reflects the higher absorption of our system when compared to a bare cell, which leads to higher radiative losses. However, while our improvement decreases with more realistic efficiency estimates, our frequency selective light trapping scheme still shows the potential to substantially improve the performance of existing semi-transparent solar cell systems.

\section{Results}
Our optimization results, shown in Figure~\ref{fig:OptimizationParameters}, find that spherical $\textrm{TiO}_2$ nanoparticles with a radius of 255 nm and an area density of 0.11 in a medium with index of refraction of 1.5 establish strong frequency selective light trapping. The scattering and absorption cross sections, seen in Figure~\ref{fig:OptimizationParameters}\color{red}a\color{black}, show desired characteristics. A broad peak in the IR spectrum provides necessary scattering for increased efficiency, while a dip in the scattering cross section around the visible spectrum maintains transparency. In addition, $\textrm{TiO}_2$ gives low absorption across all frequencies, which prevents unnecessary loss of light that would impact both efficiency and transparency.  Other materials we considered (silver, gold, and $\textrm{Ta}_2\textrm{O}_5$) have high absorptions that reduce both transparency and efficiency, resulting a low FOM. 

In addition to our calculations for a medium with index of 1.5, we find that increasing indices of refraction leads to increased performance due to increased light trapping.  We present in Table \ref{table:results} our results for an index of 1.5 to represent more realistic material properties as well as an index of 1.8 to demonstrate this increased improvement.

\begin{table}[h]
\small
	\caption{Figure of Merit Results at 60\% Transparency}
	\centering
	\renewcommand{\arraystretch}{1.5}
	\begin{tabular}{ccc}
		\hline
		Medium Index & Crystalline Silicon & Quantum Dot \\
		\hline
		1.5 & $13.3\%\pm1.3\%$ &  $2.0\%\pm1.1\%$ \\
		1.8 & $16.1\%\pm1.5\%$ & $5.6\%\pm1.2\%$ \\
		\hline
	\end{tabular}
	\label{table:results}
\end{table}

Table~\ref{table:results} shows the result of performing our FOM calculation using Shockley Queisser efficiency for each combination of active layer and medium index.  From our results we can make a couple important observations.  First, increasing the index medium from 1.5 to 1.8 increases the FOM.  For a medium index of 1.8, our optimization results find that spherical $\textrm{TiO}_2$ nanoparticles with a radius of 309nm and area density of 0.149 work best.  With a higher index, the index difference between the medium and nanoparticles decreases, leading to decreased scattering.  At the same time, light that is scattered will be more easily trapped, which offsets the decreased scattering and leads to increased absorption.  Overall, this increased absorption increases efficiency more than it decreases transparency, leading to an improved FOM.  In addition, we observe that a quantum dot active layer shows a lower FOM than silicon.\footnote[6]{A discussion of additional results obtained when considering the effects of graphene contacts can be found in Appendix \ref{app:graphene}}

Using our results for crystalline silicon in a medium of index 1.5 as the best performing, most realistic results, we determine the potential of our frequency selective light trapping model.  Such a cell gives a Shockley Queisser efficiency of $12.0\%\pm0.4\%$ with a transparency of $60.2\%\pm1.3\%$ and a 5\% haze.  This is substantially more transparent than etched solar cells and shows the potential of being more efficient than the 4\% efficiency of organic cells.  Furthermore, we only considered single material spherical nanoparticles for our optimization.  By considering multi-material, layered nanoparticles, we believe that it will be possible to obtain more effective resonances, opening up the possibility of creating far more effective semitransparent solar cells in the future.

\section{Conclusion}
We demonstrated the effectiveness of a frequency selective light trapping scheme using a nanoparticle scattering layer and photonic stack reflective layer.  We performed an optimization to determine the best scattering layer design while using a ray tracing calculation to determine a useful figure of merit.  Our results show that we can use silicon to create effective semiconductor based semitransparent solar cells with both high efficiencies and high transparencies.  Creating a 420nm thick crystalline silicon solar cells was not in the scope of this research, but some relevant experimental researches have been carried out previously \cite{thin-film1,thin-film2}.  For this paper, we only optimized for spherical, single material nanoparticles.  By performing further optimizations that account for layered nanoparticles, we believe that the performance of our light trapping scheme will improve.  In addition, with the ability to create thin enough crystalline silicon solar cells, our light trapping scheme can create semitransparent solar cells with high efficiency and visible transparency.  This frequency light trapping scheme can be used to increase the efficiency of any semitransparent solar cell, including those that use the methods described in Section \ref{sec: Background}.  This can be accomplished by replacing the active layer with an existing semi-transparent solar cell technique, such as an organic cell.

\begin{acknowledgements}
Research supported as part of the Army Research Office through the Institute for Soldier Nanotechnologies under contract no. W911NF-13-D-0001 (photon management for developing nuclear-TPV and fuel-TPV mm-scale-systems).  Also supported as part of the S3TEC, an Energy Frontier Research Center funded by the US Department of Energy under grant no.  DE-SC0001299 (for fundamental photon transport related to solar TPVs and solar-TEs).  This work was also supported in part by the MRSEC Program of the National Science Foundation under award number DMR - 1419807.  This work was also funded in part by the MIT Undergraduate Research Opportunity Program (UROP) and through a summer UROP sponsored by MIT Energy Initiative Affiliate Member Alfred Thomas Guertin PhD '60.
\end{acknowledgements}






\appendix

\section{Frequency Selective Reflector}
\label{app:reflector}

\begin{figure}[h]
	\includegraphics[width=8cm]{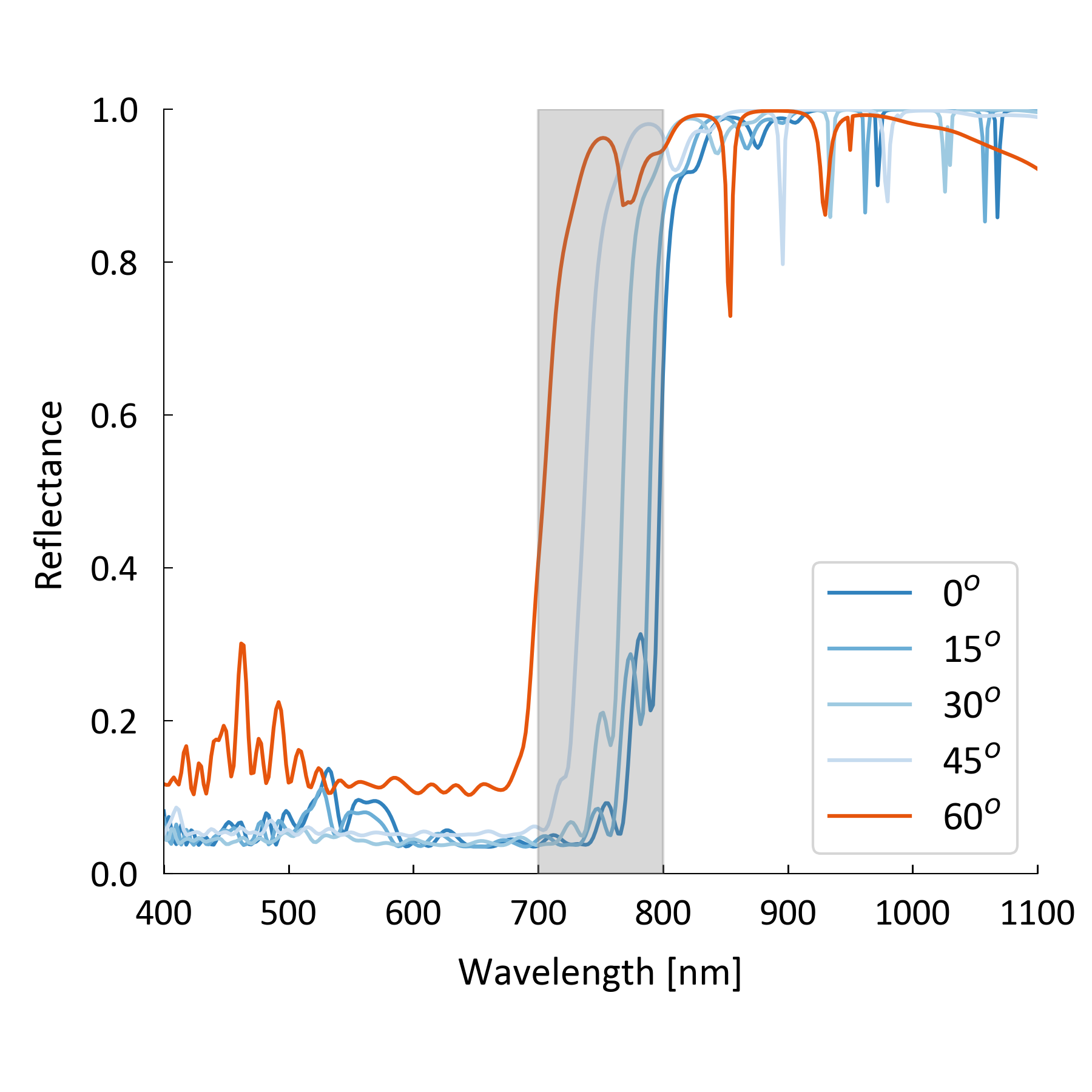}
	\caption{Reflectance of Simple Frequency Selective Filter}
	\label{fig:Ognjen}
\end{figure}

To better understand how close to ideal a frequency selective filter might be, we performed a quick optimization.  Using two materials, TiO2 and SiO2, in air, we found a multilayer structure to maximize reflection in the IR spectrum (800 - 1100nm) while minimizing reflection in the Visible spectrum (400 - 700nm), leaving the 700 - 800nm range for a transition between low and high reflectance.  In addition, the design maintains these properties for a wide range of angles.  Figure ~\ref{fig:Ognjen} shows the reflectance of the reflector design for 0, 15, 30, 45, and 60 degrees of incidence.

When averaging the reflectance for 400 - 700nm at each of the angles mentioned above, we find 5.9\%, 5.3\%, 4.3\%, 5.4\%, and 13.9\% respectively.  When averaging the reflectance for 800 - 1100nm at each angle, we find 97.8\%, 98.4\%, 98.9\%, 98.5\%, and 96.9\% respectively.  While we did not go so far as to incorporate the reflector into our Monte Carlo simulation, these numbers show very promising behavior for a simple two material design.  In addition, by separating the reflector from the rest of the solar cell by an air gap, we can use total internal reflection to trap any light outside of the escape cone of the medium material surrounding the active layer.  To improve upon this further, more materials can be used along with a figure of merit that prioritizes relevant angles.  With such factors, the behavior of a filter can be expected to improve dramatically and impact the behavior of the transparent cell minimally.

\section{Transparent Graphene Contacts}
\label{app:graphene}

For our main calculations we assume ideal transparent contacts.  For a more realistic result, we briefly consider possible contact materials.  ITO, a standard transparent conductor does not work well for our design as ITO has high absorption in the IR spectrum, and would dramatically decrease efficiency.  Rather, we must find a conductor with low absorption across a broad range of the electromagnetic spectrum.  Therefore, as a promising transparent contact, we consider using graphene which has a nearly consistent 2.3\% absorption coefficient over the frequencies we consider \cite{TransparentOrganic1,Graphene}.  To determine the effect of a graphene contact, we include an extra 2.3\% chance of absorption in our ray trace calculation every time a photon enters or exits the active layer.  We also add a 2.3\% chance of absorption to all light entering and leaving the bare cell configuration to establish an effective comparison for the FOM.

\begin{table}[h]
	\caption{Figure of Merit Results At 60\% transparency With Graphene}
	\centering
	\renewcommand{\arraystretch}{1.5}
	\begin{tabular}{>{\centering\arraybackslash}m{32mm}cc}
		\hline
		Active Layer Material & Crystalline Silicon & Quantum Dot \\
		\hline
		\hline
		Medium index of 1.5 \newline with contact loss & $7.77\%\pm1.44\%$ & $0.03\%\pm0.57\%$ \\
		\hline
		Medium index of 1.8 \newline with contact loss & $7.03\%\pm1.65\%$ & $0.42\%\pm0.68\%$\\
		\hline
	\end{tabular}
	\label{table:graphene}
\end{table}

Our ray tracing calculations, including graphene loss, give results, shown in Table~\ref{table:graphene}, reveal a dramatic decrease in our FOM. While the graphene also decreases the efficiency and transparency of the bare cell, it creates a larger effect in the light trapping scheme as individual rays pass through the graphene many times.  Additionally, the lower medium index now gives a higher FOM than the larger medium index.  This reflects the effects of a smaller escape cone. With a higher index, light that is scattered at all will be more easily trapped, leading to a higher efficiency with low contact losses. However, this also means that more IR light will be absorbed through contact loss with a higher medium than a lower medium.  As a result, the efficiency of the solar cell will decrease far more due to the graphene in our simulation with a medium of 1.8.

To minimize these losses, more work on transparent contacts is needed.  For our simulation, we assumed a uniform sheet of graphene across both sides of our active layer.  However, rather than using a sheet of graphene, a grid layout could be imagined where much of the active layer will have no contact absorption losses.  In such a case, the efficiency losses due to the addition of realistic contacts will be dramatically decreased, and might possibly become negligible.  To determine the ideal layout of graphene contacts on the surface of the active layer, further research and simulations are needed to understand the trade off between reducing charge carrier pickup and reducing contact absorption losses.

\bibliographystyle{unsrt}
\bibliography{transparent_solar_arxiv3}

\end{document}